\documentclass[sn-mathphys]{sn-jnl}
\usepackage{fix-cm}
\usepackage{comment}
\usepackage{enumitem}

\begin{document}

\title[Generative AI in Training and Coaching]{Generative AI in Training and Coaching: Redefining the Design Process of Learning Materials}


\author{\fnm{Alexander} \sur{Komar}}
\author{\fnm{Marc-André} \sur{Heidelmann}}\email{marc-andre.heidelmann@iu.org}
\author{\fnm{Kristina} \sur{Schaaff}}\email{kristina.schaaff@iu.org}

\affil{\orgdiv{IU International University of Applied Sciences}, \orgaddress{\city{Erfurt}, \country{Germany}}}


\abstract{Generative artificial intelligence (GenAI) is transforming education, redefining the role of trainers and coaches in learning environments. In our study, we explore how AI integrates into the design process of learning materials, assessing its impact on efficiency, pedagogical quality, and the evolving role of human trainers and coaches. Through qualitative interviews with professionals in education and corporate training, we identify the following key topics: trainers and coaches increasingly act as facilitators and content moderators rather than primary creators, efficiency gains allow for a stronger strategic focus but at the same time the new tools require new skills. Additionally, we analyze how the anthropomorphism of AI shapes user trust and expectations. From these insights, we derive how tools based on GenAI can successfully be implemented for trainers and coaches on an individual, organizational, systemic, and strategic level.}




\maketitle
\section{Introduction}\label{sec:introduction}

The rapid development of generative AI (GenAI) has significantly impacted education, especially in the creation of learning materials \cite{Bearman2023}. Tools based on GenAI, can help generate content, assist in lesson planning, and personalize learning experiences \cite{Farrelly2023}. These technological advancements raise important questions about the role of GenAI in educational scenarios, the reliability of AI-generated materials, and the level of human oversight required.
However, compared to institutional learning, the largely unregulated and market-driven field of coaching and training remains significantly under-researched  \cite{Heidelmann2022}. Given the rise of GenAI, engaging with this field of practice is particularly relevant -- especially concerning the quality of AI-generated materials, the extent to which trainers and coaches should rely on GenAI, and the level of trust in AI-generated content. Moreover, anthropomorphism, i.e. attributing human-like characteristics to non-human things, presents new challenges in user interaction, acceptance, and potential overreliance \cite{Schaaff2024}. We emphasize the necessity for AI literacy among trainers and coaches to ensure that AI-generated materials align with educational goals rather than merely automating content generation.

In our study, we will address the following research questions:

\begin{description}[leftmargin=1cm]  
    \item[\textbf{RQ1:}] How does the use of GenAI tools change the role of trainers and coaches in the design process of learning materials?
    \item[\textbf{RQ2:}] What advantages and challenges do trainers and coaches perceive in the use of GenAI tools?
    \item[\textbf{RQ3:}] What new tasks and competencies emerge for trainers and coaches due to the use of AI?
    \item[\textbf{RQ4:}] How does the anthropomorphism of AI influence the interaction of trainers and coaches with GenAI and the trust in these tools?
\end{description}

Research on the use of GenAI in coaching and training is particularly relevant due to this field's low institutionalization and academic integration. This is largely due to a lack of legal standards and established academic disciplines~\cite{Heidelmann2022}. Consequently, the quality of coaching and training remains highly heterogeneous. Unlike institutional education settings, the use of GenAI in this field is subject to even fewer regulations. This makes it crucial to examine how GenAI is implemented in practice and how it impacts coaching and training processes.

Our study extends existing research by demonstrating how GenAI is not merely a tool for efficiency but a transformative force reshaping the role of trainers and coaches. While previous studies \cite{neumann2024, Hein2024} emphasize the ability of GenAI to automate content creation, our findings highlight a shift from content generation to quality assurance. Trainers increasingly act as curators, reviewing and refining AI-generated materials rather than creating content from scratch. Moreover, to the best of our knowledge, we are the first to explore the anthropomorphism of GenAI in the design of learning materials. Unlike prior research focusing on AI adoption \cite{watanabe2023}, our findings show that trainers often perceive GenAI as a creative partner, influencing their prompting behavior and workflow. 
Finally, our findings stress the necessity of human oversight. GenAI should be seen as an assistive -- not a substitution -- tool in instructional design. 
\\\\
Consequently, our contributions are as follows: 
\begin{itemize}
    \item We provide empirical evidence that GenAI shifts the role of trainers and coaches from content creators to quality assessors. 
    \item We highlight the growing need for structured AI literacy for trainers and coaches, as many trainers rely on informal learning strategies.
    \item We explore the impact of anthropomorphism of trainers and coaches when creating learning materials. 
\end{itemize}
By focusing on the minimally institutionalized and largely academically detached field of coaching and training, we address an important research gap.
\section{Theoretical Background}

Integrating GenAI in training and coaching requires a clear conceptual understanding of these domains and the instructional design frameworks that support them. This section defines training and coaching and introduces the ADDIE model \cite{Kerres2018} as a fundamental instructional design approach. As our research focuses on the design process of learning materials, we put our focus on the design phase of the ADDIE model. 

\subsection{Training and Coaching}

Training and coaching are broad and often ambiguous terms, encompassing diverse concepts depending on the theoretical origin and professional theoretical positioning \cite{Heidelmann2022}. Both are distinct but also complementary approaches to educational and professional development. Training is typically structured and aims at imparting specific knowledge, skills, or competencies to learners. It follows predefined curricula and measurable learning objectives \cite{Becher2022}. Coaching, on the other hand, is a more individualized, goal-oriented process focusing on personal and professional development. It involves guiding learners through self-reflection, problem-solving, and skill enhancement rather than providing direct instruction \cite{Becher2022}.

\subsection{The Design Phase in the Didactic Work Process}
The design of didactic materials plays a crucial role in learning scenarios. Therefore, it is important to follow a structured approach. A widely used instructional design model in media-didactic planning is ADDIE (Analysis, Design, Development, Implementation, and Evaluation)  \cite{Kerres2018}. The design phase, central to this study, structures learning content and defines pedagogical strategies. It serves as an interface between the analysis phase, in which requirements and learning objectives are formulated, and the subsequent development phase, in which the specifically planned content and materials are created  \cite{Kerres2018}. AI integration in the design phase can enhance efficiency but also requires careful quality control.

\section{Use of GenAI for Training and Coaching}
AI has become a powerful tool in education, offering innovative ways to enhance learning experiences \cite{Bearman2023, Popenici2017}. Prior research has highlighted the growing influence of GenAI in education \cite{Villareal2023, Perera2023, zouhaier2023impact}. For instance, AI-based chatbots can be used for personalized learning support \cite{Wollny2021, okonkwo2021, timo25}. Still, AI can help to automate tasks -- be it administrative \cite{okonkwo2021} or organizational \cite{mueller2025} -- personalize learning pathways \cite{Borah2024}, and enhance instructional design \cite{zawacki-richter2019}. GenAI applications like ChatGPT have demonstrated their ability to create structured lesson plans, provide immediate learner feedback \cite{latif2023}, and generate interactive learning content. 
Studies on the impact of GenAI on coaching and professional training suggest a shift toward blended intelligence, where GenAI serves as an assistant in content generation while human trainers refine and contextualize the material to align with pedagogical goals~\cite{holmes2019artificial, passmore2023}. The role of coaches shifts from direct knowledge transmission to moderating and guiding digitally supported learning processes, a development that \cite{Guilherme2019} describes as `learnification'.
Furthermore, research on anthropomorphism of GenAI indicates that human-like interactions can increase trust but also introduce risks related to over-reliance on AI-generated content \cite{nass2000}. 

However, despite these advantages, challenges remain, particularly concerning the accuracy and quality of AI-generated content \cite{bender2021}. GenAI may produce misleading or incorrect information \cite{rawte2024}, which can compromise the reliability of educational content. Trainers and coaches must therefore critically evaluate and refine AI-generated materials to ensure factual accuracy and pedagogical effectiveness \cite{neumann2024}.
Moreover, the ethical implications of GenAI in education -- such as bias in AI-generated content -- must be considered \cite{busker23}. The underlying models learn from vast datasets that may contain social biases, potentially perpetuating stereotypes or inaccurate representations in generated learning materials. As a result, trainers and coaches need to approach AI-generated content critically and ensure that it aligns with pedagogical goals and ethical considerations.

\section{Experimental Setup}
\label{sec:experimentalsetup}
In our study we, used a qualitative research design, utilizing semi-structured interviews with five professional trainers and coaches who actively incorporate GenAI into instructional design. In the following, we will describe our experimental setup.

\subsection{Participant Selection}

For our study, we conducted interviews with trainers and coaches from Germany working in formal education, professional development, and individual coaching. Their core industries included the automotive sector, coaching, and continuing education. This diversity provided insight into different facets of GenAI use and how professionals engage with the technology.

We focused on individuals actively involved in developing learning materials and assessing the impact of AI tools on their work. Participants were selected for their expertise in using GenAI for content creation and learner engagement. The generated content includes images, infographics, texts, and videos. To capture a range of perspectives, we included both experienced users and those newly exploring GenAI, highlighting varied approaches to adopting this evolving technology.

Table \ref{tab:participants} provides an overview of the interviewed participants, including their roles, years of experience, and the GenAI tools they use in their work.

\begin{table}[htp]
\centering
\begin{tabular}{@{}llp{7.7cm}@{}}
\toprule
\textbf{Participant} & \textbf{Experience} & \textbf{AI Tools Used} \\
\midrule
P1 (Trainer) & 8 years  & ChatGPT, MS Copilot \\
P2 (Trainer) & 6 years  & ChatGPT, Synthesia (passive) \\
P3 (Coach)   & 11 years & ChatGPT \\
P4 (Trainer) & 7 years  & ChatGPT, Consensus, Perplexity, Midjourney, Adobe Firefly, ElevenLabs \\
P5 (Trainer) & 16 years & ChatGPT, Stable Diffusion \\
\bottomrule
\end{tabular}
\caption{Overview of Participants}
\label{tab:participants}
\end{table}

\subsection{Data Collection and Analysis}

The primary data collection method consisted of semi-structured interviews conducted via online video conferencing. The guideline for the interviews was developed based on the research questions introduced in Section \ref{sec:introduction}. When developing the guidelines, particular care was taken to formulate neutral and open questions. Due to its semi-structured nature, the guide provides a structured and at the same time flexible basis for recording both explicit and implicit experiences and perceptions of the participants. Each interview lasted approximately 45--60 minutes and followed a structured guideline. The interviews were recorded with participant consent and subsequently transcribed for analysis.\footnote{The full interview guideline can be found at \url{https://github.com/iu-ai-research/GenAI-in-Training-Coaching}.}

We analyzed the participants' responses for their direct content using qualitative content analysis, following \cite{Mayring2012,Selvi2019} and examined overarching themes and connections \cite{Selvi2019}, thereby capturing their subjective meanings and experiential horizons.

Our analysis follows both deductive and inductive approaches: Deductively, categories were derived from theoretical prior knowledge and existing hypotheses regarding the impact of GenAI tools on trainers and coaches and then applied to the material \cite{Mayring2012, Selvi2019}. At the same time, the method allowed for an inductive expansion of the categories, enabling the consideration of new insights emerging from the data.
Following this approach, this combination of deductive categorization and inductive refinement, existing assumptions could be tested, and new perspectives could be gained from the data.

\begin{figure}[htp]
    \centering
    \includegraphics[width=0.9\linewidth]{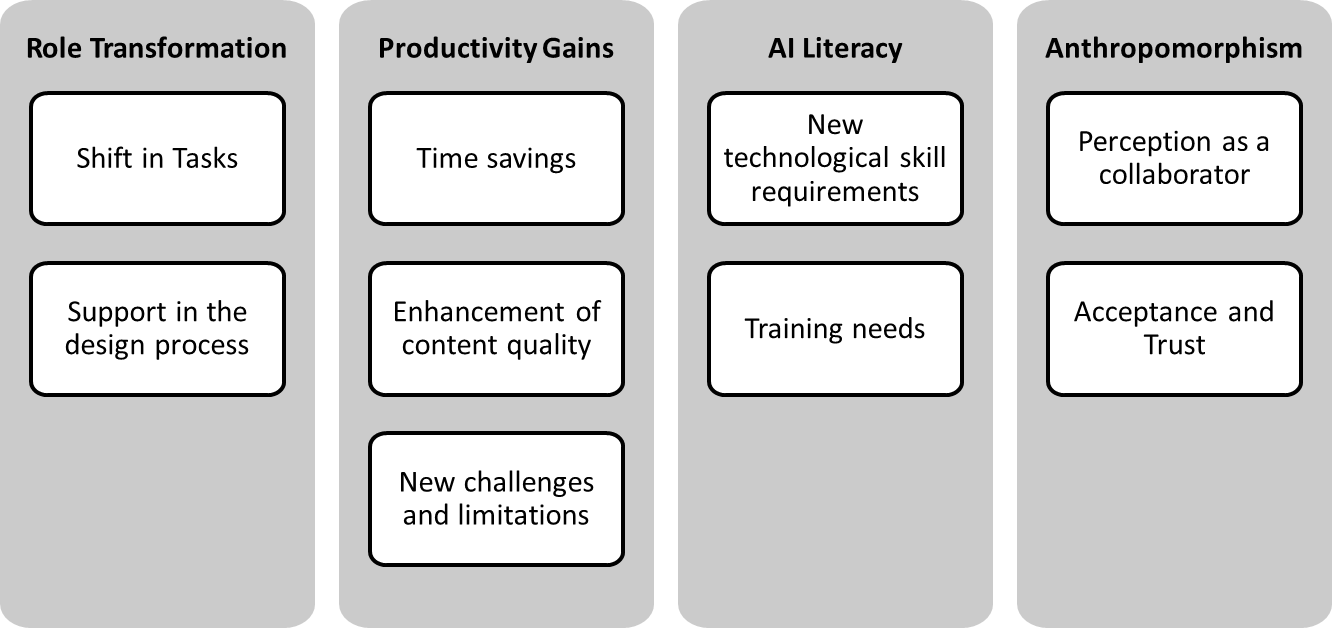}
    \caption{Main Categories and Sub-Categories for Qualitative Analysis}
    \label{fig:categories}
\end{figure}

From the coding process, we developed the following main categories for qualitative analysis: \textit{Role transformation}, \textit{productivity gains}, \textit{AI literacy}, and \textit{anthropomorphism}. The main categories including the sub-categories we derived for each category are illustrated in Figure \ref{fig:categories}. 

The subsequent qualitative analysis of the interviews will present the findings and assign them to the corresponding categories and subcategories.

\section{Results of the Qualitative Data Analysis}
In the following, we will present the results of our qualitative data analysis, structured according to the developed categories. Each section highlights key themes that emerged from our interviews and directly addresses the research questions.
As described in Section \ref{sec:experimentalsetup}, we analyzed the transcribed interviews using qualitative content analysis~\cite{Mayring2012, Selvi2019}. We applied thematic coding to identify recurring patterns and emergent themes relevant to the impact of GenAI on the design process of learning materials. 
\subsection{Role Transformation}
We will start by examining the potential influences of integrating GenAI on the changing role of trainers and coaches in the design process of learning materials (\textit{RQ1}). We analyze how the introduction of GenAI tools impacts their workflow and task distribution.

\subsubsection{Shift in Tasks}

The integration of GenAI in training and coaching shifts traditional responsibilities from manual content creation toward quality control and optimization. P1 highlighted how GenAI tools reduce the effort required for information retrieval and material preparation, allowing more time for complex tasks. Similarly, P2 described automation as an opportunity to focus on creative and strategic work, noting that trainers are increasingly assuming the role of reviewers while GenAI takes over creator tasks: \textit{``I go into the reviewer mindset, I don't go into the creator mindset [...]''}\footnote{All quotations from the German interviews have been translated into English by the authors.}.

P4 emphasized that GenAI allows for greater learner engagement by reducing time spent on routine work, enabling a more personalized approach. While P5 acknowledged this trend, they also pointed out that full automation is only feasible with highly specialized GenAI. Certain tasks, particularly those involving sensitive data, still require manual oversight.

The interviews indicate that GenAI automates repetitive tasks but expands the role of trainers to include more strategic and interpersonal activities. The challenge lies in using GenAI purposefully to ensure high-quality learning content and individualized learning support.

\subsubsection{Support in the Design Process}

The participants observed that GenAI systems increasingly support content creation and are perceived as creative assistants. P1 noted that GenAI offers diverse perspectives and stimulates creative processes, making it more than just an automation tool—it serves as a co-creator: \textit{``[...] to get support from different directions and generate new ideas'}'.

P2 emphasized the advantage of GenAI in automating standardized tasks such as structuring learning modules, and freeing time for creative and strategic work. Similarly, P4 noted efficiency improvements through pre-designed prompts, allowing for a stronger focus on individualized and complex topics.

A particularly interesting insight is the perception of GenAI as a discussion partner. P5 described the role of GenAI in defining learning objectives and structuring content, which makes material development more efficient. GenAI also acts as a sparring partner in the conception phase, which facilitates collaborative brainstorming processes.

These findings highlight that GenAI not only facilitates routine work but also enhances creative processes and collaboration in instructional design.

\subsection{Productivity Gains} \label{sec:productivity}

To answer \textit{RQ2}, we will analyze the impact of integrating GenAI tools on the working methods of trainers and coaches, as well as the influence on efficiency and the quality of their outcomes in the design process. 

\subsubsection{Time Savings} \label{sec:timesavings}

The integration of GenAI significantly reduces the time required for learning material creation. Trainers and coaches benefit primarily in terms of rapid information retrieval and text generation. P1 emphasized that GenAI accelerates manual research, while P3 noted that GenAI tools compensate for writing weaknesses and automate time-consuming tasks such as calculating key figures: \textit{``So the advantage for me personally is also this time saving''}.

P4 highlighted the efficiency of GenAI in quickly generating learning objectives and structured content. P5 described how AI-generated drafts streamline the design process. However, some limitations remain: P4 pointed out that in certain cases, GenAI implementation is still too complex to achieve actual efficiency gains.
The findings suggest that GenAI tools often provide substantial workload reductions, particularly in well-structured tasks. However, the varying degrees of effectiveness indicate that GenAI must be selectively integrated and further developed to reach its full potential.

\subsubsection{Enhancement of Quality} \label{sec:quality}

GenAI improves the quality of coaching and training materials by enhancing structure and professional presentation. P1 noted that GenAI-produced materials are more appealing and better structured, simplifying training preparation: \textit{``My personal feeling is that it makes the whole thing look more professional''}.

P2 acknowledged that while current GenAI tools are not yet fully mature, they enhance consistency and continuity, promoting professionalism and reliability. P4 emphasized the role of GenAI in stimulating creativity and generating ideas that might otherwise go unnoticed.
P5 highlighted that the usefulness of GenAI depends on its training and application context—while it accelerates some processes, it remains inefficient in others.

The interviews confirm the potential of GenAI to enhance material quality, although further improvements are necessary to meet all didactic requirements. GenAI helps trainers and coaches professionalize materials, inspire creativity, and improve preparation, but it is not universally applicable.

\subsubsection{New Challenges and Limitations}\label{sec:challenges}

Despite the numerous advantages, the interview analysis also reveals challenges. Advantages include content structuring, efficiency gains, and creative stimulation. However, P1 stressed that while GenAI enhances material professionalism, P4 pointed out the necessity of quality control, as GenAI struggles with complex or context-dependent tasks.

P3 noted that adapting content generated by AI to personal styles can be time-consuming, limiting automation’s benefits. In some cases, they preferred manual content creation. P5 criticized the lack of GenAI of innovation and emotional intelligence, which can hinder its acceptance in emotionally driven learning processes. Moreover, P5 mentions that it is not always possible to rely on the output. 

Ethical concerns involve data privacy and failure of GenAI to support gender-inclusive language, as noted by P4. Additionally, excessive GenAI dependence could undermine creativity and trainer autonomy.

The findings illustrate that GenAI can be a powerful tool but must be used thoughtfully. Human competencies such as creativity, emotional intelligence, and innovation remain irreplaceable, requiring a balanced human-AI collaboration.

\subsection{AI Literacy}

The introduction of GenAI in the field of further education requires new competencies from trainers and coaches and what training strategies are necessary to meet these demands which is reflected in \textit{RQ3}. The qualitative analysis of interviews highlights various aspects that illustrate how professionals need to adapt to the rapidly evolving technology.

\subsubsection{New Competency Requirements} \label{sec:competency}

Our results show, that the use of GenAI in instructional design requires new skills from trainers and coaches, particularly in prompting, technological adaptability, and critical evaluation of AI-generated content: 
\begin{itemize}
    \item \textit{Prompting}: P1 and P3 stressed that precise prompt formulation is essential for quality GenAI output. P4 described learning advanced techniques to maximize potential of GenAI.
    \item \textit{Technological Adaptability}: P5 noted the need to stay up to date with GenAI developments and test new tools regularly, which can sometimes be overwhelming. P1 added that understanding interactions between different GenAI tools is crucial for effective use.
    \item \textit{Critical Understanding}: P3 emphasized that GenAI outputs are not always perfect, requiring professionals to understand GenAI mechanics to refine outputs and recognize limitations.
\end{itemize}

The interviews indicate that continuous learning and adaptability are critical for leveraging GenAI in education effectively. Professionals must expand their competencies to remain effective in an increasingly digitalized learning environment.

\subsubsection{Training Needs} \label{sec:training}

The adoption of GenAI in learning material development necessitates ongoing education for trainers and coaches. However, the interviews suggest that existing training programs often fail to meet specific needs, leading professionals to rely on informal and self-directed learning.

P1 reported that internal training sessions were not useful, as they lacked relevance to practical applications: \textit{``We had various internal initiatives, including training programs, in other words, smaller things, none of which I thought were good''}. P2 preferred hands-on learning via LinkedIn and similar platforms. Regarding informal learning, 
P3 described developing GenAI competencies through experimentation, online resources, and social media insights. P5 confirmed that tutorials, peer discussions, and specialized training programs enhance their GenAI skills.
P4 emphasized that combining theoretical knowledge with hands-on practice, such as case studies and GenAI tool discussions, is an effective way to deepen the own expertise.

These findings suggest that flexible, application-oriented learning formats combining theory and practice are essential for competency development. Many professionals prefer self-directed learning to keep pace with rapid technological advancements and maximize the potential of GenAI.

\subsection{Anthropomorphism}
The anthropomorphism of GenAI affects the perception of trainers and coaches and their use of this technology. To answer \textit{RQ4}, we will highlight the key topics from the qualitative analysis of interviews that shape interactions between humans and machines.

\subsubsection{Perception as A Collaborative Partner}\label{sec:collab}

Trainers and coaches increasingly view GenAI as a creative partner that supports and enhances learning material development. The interviews highlight the role of GenAI beyond automation, demonstrating its ability to inspire and organize thoughts.

P3 described GenAI as a brainstorming partner that provides suggestions for evaluation and refinement: \textit{``But I have a brainstorming partner, I would say, who I can somehow judge for myself''}. P4 noted that GenAI compensates for the lack of peer discussions by structuring ideas and offering new perspectives.
If it comes to anthropomorphism and partnership, P4 found GenAI interaction to be collaborative and human-like, fostering trust and acceptance. P5 saw GenAI tools as buddies or tutors that actively contribute to the learning process.

Our interviews confirm that GenAI is perceived as more than a tool—it is an interactive and dynamic companion that enhances creativity and instructional structuring.

\subsubsection{Trust and Acceptance}\label{sec:trust}

The perception and acceptance of anthropomorphized GenAI tools vary greatly among coaches and trainers. While some perceive the human-like interaction as helpful and trust-enhancing, others prefer a clear distinction between humans and machines.

P1 finds the human traits of GenAI pleasant, as they make interactions more natural and trustworthy. P2 appreciates the ability to communicate with GenAI as if it were a human, which simplifies and enhances communication efficiency. P4 emphasizes that the simple and dialogical interaction without technical expertise facilitates use and increases acceptance: \textit{``"You don't have to memorize any coding language or anything''}.

P3 remains neutral to negative toward anthropomorphism and prefers a clear separation to view the technology realistically. P5 perceives excessive anthropomorphism as exaggerated and prefers to see GenAI as a tool rather than a partner, rejecting any emotional attachment to GenAI.

The interviews highlight the necessity of a balanced approach that offers users flexible interaction options. Striking the right balance between human-like interaction and functionality could improve the acceptance of GenAI in the design process by appealing to both technology-savvy and skeptical users.

\subsubsection{Exploratory Study: Influence of Anthropomorphism on Prompting}

To explore the interaction with GenAI tools, we conducted an additional exploratory study, where participants were asked: \textit{``How would you formulate a prompt to create an image of a flipchart with the inscription [text]?''}

The analysis of the resulting prompts shows that the degree of anthropomorphism of GenAI significantly influences the formulation of prompts and task distribution:

\begin{itemize}
    \item \textit{Low Anthropomorphism}: P1 and P5 perceive GenAI as a neutral tool. P1 formulates clear, precise instructions without creative expectations. P5 uses distant terms like `thing' and expects only the fulfillment of well-defined functions but still incorporates minimal human-like politeness, such as `please'.
    \item \textit{Moderate Anthropomorphism}: P2 and P4 use more polite and slightly open-ended formulations. P2 trusts GenAI to aesthetically design visual aspects, while P4 provides detailed yet clearly defined instructions that limit the creative freedom of GenAI.
    \item \textit{High Anthropomorphism}: P3 allows GenAI creative freedom and expects it to act interpretatively, such as by emphasizing attention-grabbing elements. This demonstrates a high level of trust in the capabilities of GenAI and implicitly assigns it human-like attributes.
\end{itemize}

In summary, a higher degree of Anthropomorphism leads to more open and less precise prompts, allowing GenAI greater creative freedom, whereas lower anthropomorphism results in clear and functional instructions. The perception of GenAI as either a tool or a creative partner significantly influences interaction styles and task distribution.

\section{Discussion of the Results}
We will now discuss the key findings of the qualitative content analysis and link them to the context of existing research. Our goal is to examine and evaluate the assumptions formulated at the beginning of the study based on the collected data. To achieve this, insights gained from the interviews are placed in the context of existing academic literature and aligned with theoretical foundations.
Our analysis focuses on the evolving role of trainers and coaches in the design process due to the use of GenAI tools, efficiency and quality gains in the development of learning materials, and the challenges arising from new competency requirements and the anthropomorphism of GenAI. These four key aspects were extensively discussed in the interviews and form the core of this study.

\subsection{Role Transformation: The Changing Role of Trainers and Educators}

The qualitative analysis demonstrates that the role of trainers and coaches is undergoing a profound transformation due to the use of GenAI tools (\textit{RQ1}). These technologies are increasingly taking over tasks that were previously carried out exclusively by humans.

This includes the automatic creation of learning materials and exercises, which were traditionally developed manually by professionals \cite{neumann2024}. The automation of repetitive and standardized tasks by GenAI allows trainers to focus on more creative and strategic activities, ultimately enhancing their efficiency~\cite{neumann2024}. The qualitative analysis of the interviews reveals that trainers increasingly perceive their role as reviewers rather than creators. This shift highlights the restructuring of their tasks towards reviewing and optimizing AI-generated content.

Another key insight from the interviews is the perception of GenAI as a creative partner in idea generation and concept development. This is in line with the findings by \cite{Hein2024} who point out that GenAI can be used to analyze vast amounts of data and generate personalized content, which can then be further refined by subject matter experts.

Moreover, the interviews confirm the findings by \cite{watanabe2023} who state that the acceptance of GenAI is closely linked to their perceived usefulness. Therefore, it is crucial that trainers and coaches view GenAI tools as a support rather than a burden. 

\subsection{Productivity Gains: Efficiency vs. Quality Trade-Offs}

The automation of time-intensive processes by GenAI represents a key potential benefit (\textit{RQ2}). For example, GenAI can automatically generate learning materials and exercises, resulting in significant time savings for trainers and coaches \cite{zawacki-richter2019}. These efficiency gains enable professionals to invest more time in individualized learner support instead of focusing on administrative tasks. This is also confirmed by our results presented in Section \ref{sec:timesavings}. 

Beyond improving efficiency, GenAI can also enhance the quality of learning materials. According to \cite{neumann2024}, GenAI can personalize learning content and tailor it to the needs of learners. Our results presented in Section \ref{sec:quality} confirm this observation. 

However, the qualitative analysis also highlights the limitations of this technology. Participants emphasized that AI-generated content often requires post-processing to meet the desired didactic standards. Similarly, \cite{Hein2024} stress that monitoring and adjusting generated content is essential for quality assurance, underscoring the critical role of human oversight in the quality control process.

Another challenge is the adaptability of GenAI to individual styles and emotional needs. AI-generated content often does not align with the personal style of subject matter experts and must therefore be revised. This issue highlights that, despite efficiency gains, human experts continue to play a crucial role in fine-tuning and ensuring the quality of learning materials \cite{Krupp2023}.

In addition to qualitative and didactic challenges, ethical and data privacy concerns add further complexity to the use of GenAI in further education. These concerns necessitate a conscious and reflective approach to ensure that both ethical standards and quality requirements are met~\cite{Slimi2023}.

\subsection{AI Literacy: Future Competency Requirements}

The precise formulation of prompts is highlighted as a key skill. \cite{Hein2024} emphasize that the quality of AI-generated content largely depends on how accurately and purposefully instructions are formulated. This competency is crucial for training professionals to achieve the desired results and effectively integrate GenAI into their workflows. The results illustrated in Section \ref{sec:training} confirm the importance of new skills like efficient prompting techniques (\textit{RQ3}). 

Furthermore, the qualitative analysis presented in Section \ref{sec:competency} reveals that the rapid technological development in the field of GenAI necessitates continuous learning for trainers and coaches. One-time training sessions are insufficient; instead, an ongoing learning process is required to keep up with the latest advancements and continuously update skills. This suggests that practical, self-directed learning plays a crucial role in competency development which is also in line with the findings presented in \cite{vladova2023}.

Our analysis also highlights a significant demand for further training that is not always met by existing programs. As a result, many trainers and coaches prefer informal and individualized learning resources to acquire relevant competencies. This preference indicates a strong inclination toward practical and application-oriented learning approaches that align with the specific work requirements of professionals. Training formats that effectively combine theory and practice appear to be particularly successful.

In addition to technological competencies, critical and subject-specific skills remain essential. The ability to meaningfully structure AI-generated content from a didactic perspective and integrate it into the learning process remains a core responsibility of trainers and coaches. Only by the individual adaptation of materials can trainers ensure that learning objectives are effectively met \cite{Hein2024}.

\subsection{Anthropomorphism: Impact on Trust and User Interaction}

Our qualitative analysis revealed that trainers and coaches perceive GenAI not only as technical tools but as partners in the design process (\textit{RQ4}). They describe GenAI tools as supportive and inspiring, particularly in structuring thoughts and developing new ideas. This perception is reinforced by the anthropomorphic traits of GenAI, which foster trust in the technology and, according to~\cite{Kim2023}, increase professionals' willingness to efficiently integrate these tools into their work.
Trainers and coaches benefit from GenAI acting as a natural conversational partner, which is especially advantageous in online coaching scenarios where personal interaction is not always possible \cite{stenzel2024}.

However, as illustrated in Section \ref{sec:trust} some participants expressed concerns that excessive anthropomorphism of GenAI could lead to unrealistic expectations of its capabilities. When these expectations are not met, frustration may arise. This discomfort is linked to the `Uncanny Valley' phenomenon, which occurs when GenAI appears human-like but still feels artificial \cite{Kim2024}. This discrepancy can create uncertainty and distrust, potentially limiting the acceptance of GenAI in coaching and training contexts.

Additionally, there is a risk that trainers and coaches, due to excessive trust in the seemingly competent and approachable nature of GenAI, may lose control over the qualitative process of material creation. Over-reliance on GenAI tools and uncritical adoption of their outputs can be problematic. According to \cite{rawte2024}, a major risk are hallucinations -- situations where GenAI generates responses that appear correct but, upon closer examination, are found to be factually inaccurate or erroneous. Uncritical use of such content can lead to quality loss and misinformation among participants and coaches.

\section{Conclusion And Future Work}

The findings of our study indicate that GenAI offers considerable potential for enhancing both efficiency and quality in professional development. However, to fully use its potential requires deliberate application and practice-oriented training. Trainers and coaches, as early adopters, increasingly assume hybrid roles that integrate technological and pedagogical competencies. Our findings address the research gap that the use of AI in the largely non-institutionalized and predominantly privately organized field of professional development has thus far remained underexplored. By focusing on the design of learning materials, we demonstrate how these actors function as knowledge intermediaries, transferring insights from practice back into formal educational contexts.

In \textit{RQ1} we wanted to determine how the use of GenAI tools changes the role of trainers and coaches in the design process of learning materials. 
Our results show that the role of trainers and coaches in the design process of learning materials has undergone significant changes due to the integration of GenAI. 

In \textit{RQ2} we addressed new tasks and competencies emerging for trainers and coaches through the use of GenAI. We found that GenAI increasingly automates repetitive tasks such as the creation of materials and exercises, while trainers and coaches are shifting towards roles focused on quality assurance and facilitation. This shift enables efficiency gains, allowing professionals to allocate more time to strategic, creative, and individualized tasks.

In our research, in \textit{RQ3} we aimed to understand what advantages and challenges trainers and coaches perceived in using GenAI tools. 
Our findings show that using GenAI allows trainers and coaches to primarily focus on the didactic adaptation and quality control of AI-generated content, while direct content creation decreases. They act as curators and facilitators, leveraging the potential of GenAI in a targeted manner.
Moreover, GenAI is increasingly perceived as a creative partner that provides inspiration and fosters new ideas. This collaboration reshapes the relationship between humans and machines, as GenAI becomes an active supporter in the design process.
Task automation allows trainers to allocate more time to individualized support and strategic planning, thereby enhancing the quality and relevance of learning processes.

In \textit{RQ4} we focused on the question of how the anthropomorphism of GenAI influences the interaction of trainers and coaches with these tools as well as their trust in them.

As a result, we found that the anthropomorphism of GenAI is viewed ambivalently. While it enhances acceptance and promotes the perception of GenAI as a partner, there is also a risk of uncritical trust, potentially limiting the autonomy and creativity of professionals.

Overall, our results show that the use of GenAI presents both potential and challenges on various levels. 
On an \textit{individual} level, the evolving role requires additional competencies, particularly in didactic evaluation and the effective use of GenAI tools. On an \textit{organizational} level, educational institutions and companies should provide high-quality, practice-oriented training that integrates theory and application. Professional networks, tutorials, and communities support independent learning and adaptability. From a \textit{systemic} perspective, ethical and data protection standards must be established through transparent guidelines and targeted training for trainers and coaches to ensure the safe and responsible use of GenAI while safeguarding learner privacy. \textit{Strategically}, GenAI should be leveraged as a supportive tool that automates repetitive tasks while preserving human expertise, with customizable interaction modes fostering trust without compromising professional autonomy.

Further research is needed to determine the optimal balance between human-like GenAI interaction and maintaining professional autonomy. 
Future research should also focus on the long-term effects of GenAI integration in training and coaching. As technologies based on GenAI continue to evolve, it is essential to continuously monitor the dynamic interaction between coaches, trainers, learners, and GenAI. In particular, longitudinal studies should be conducted to examine how the role of trainers and coaches changes over time and what new competency requirements emerge as a result of technological advancements.
On an ethical and systemic level, another important research area should be the in-depth investigation of the ethical implications of GenAI use in the field of continuing education, particularly concerning the trustworthiness and accountability of AI-generated content.
Therefore, their practical handling of GenAI is particularly significant for scientific innovations in this field. Future research should place greater emphasis on this aspect, rather than solely focusing on the use of GenAI in institutionalized educational settings, such as schools and universities.


\bibliography{20-bibliography}


\end{document}